# Locating the Atoms at the Hard-soft Interface of Gold Nanoparticles

Weilun Li[1], Bryan D. Esser[2], Wenming Tong[3], Anchal Yadav[4,5], Colin Ophus[6], Changlin Zheng[7], Scott D. Findlay[1], Timothy Petersen[2], Alison M. Funston[4,5]*, Joanne Etheridge[1,2,8]*

[1]School of Physics and Astronomy, Monash University, VIC, 3800, Australia

[2]Monash Centre for Electron Microscopy, Monash University, VIC, 3800, Australia

[3]School of Biological and Chemical Sciences and Energy Research Centre, Ryan Institute, University of Galway, H91 TK33, Galway, Ireland

[4]School of Chemistry, Monash University, VIC, 3800, Australia

[5]ARC Centre of Excellence in Exciton Science, Monash University, VIC, 3800, Australia

[6]National Centre for Electron Microscopy, Molecular Foundry, Lawrence Berkeley National Laboratory, Berkeley, CA, United States of America

[7]State Key Laboratory of Surface Physics and Department of Physics, Fudan University, Shanghai 200433, China

[8]Department of Materials Science and Engineering, Monash University, VIC, 3800, Australia

*e-mail:
alison.funston@monash.edu
joanne.etheridge@monash.edu

**Abstract**

Surface structure affects the growth, shape and properties of nanoparticles. In wet chemical syntheses, metal additives and surfactants are used to modify surfaces and guide nanocrystal growth. To understand this process, it is critical to understand how the surface structure is modified. However, measuring the type and arrangement of atoms at hard-soft interfaces on nanoscale surfaces, especially in the presence of surfactants, is extremely challenging. Here, we determine the atomic structure of the hard-soft interface in a metallic nanoparticle by developing low-dose imaging conditions in four-dimensional scanning transmission electron microscopy that are preferentially sensitive to surface adatoms. By revealing experimentally the copper additives and bromide surfactant counterion at the surface of a gold nanocuboid and quantifying their interatomic distances, our direct, low-dose imaging method provides atomic-level understanding of chemically sophisticated nanomaterial surface structures. These measurements of the atomic structure of the hard-soft interface provide the information necessary to understand and quantify surface chemistries and energies and their pivotal role in nanocrystal growth.

**Introduction**

Just over two decades ago, a breakthrough in wet chemical synthesis enabled the growth of metal nanocrystals into selected shapes [1,2]. This breakthrough has launched exciting applications in nanophotonics, sensing, catalysis, optoelectronics and biomedicine, which seek to optimize the properties of nanocrystals by tuning their shape, size and/or facet crystallography [3–6]. However, despite over two decades of intense research, key questions regarding the fundamental mechanisms controlling metal nanocrystal growth and morphology remain.

Pivotal to nanocrystal growth and shape control is the modification of the energy of surface facets using surfactants and/or absorbates (such as ions and metal additives), thereby controlling the rate at which different facets grow and directing the final nanocrystal morphology [7–10]. For example, single crystal cuboctahedra gold seeds grown with the cationic surfactant cetyltrimethylammonium bromide (CTAB) can form spheres, single crystal nanorods or nanocuboids, depending on whether they are in the presence of no additive, silver ($Ag^+$) or copper ($Cu^{2+}$), respectively [11–17]. Remarkably, these metal adsorbates are essential to induce symmetry breaking of the initial seed particle [12,18] and thereby facilitate the final anisotropic shape.

To understand such processes, and more broadly to understand how a nanoparticle interacts with its environment, it is critical to understand the atomic structure at the nanocrystal-metal additive-surfactant interfaces, in concert with the overall nanoparticle size, shape and surfactant chemistry. However, here lies a fundamental problem: nanocrystal facets are typically only a few atoms wide and are decorated with surfactant and/or trace amounts of metal additives. Measuring the atomic structure – that is, the type, position and relative arrangement of atoms – at the nanofacet-metal adsorbates-surfactant interface remains extremely challenging [19]. Methods with both sufficient resolution and sensitivity to detect a monolayer (or sub-monolayer) of ions *simultaneously* with the nanoparticle surface and the surfactant have been elusive.

Advanced characterizations, such as X-ray diffraction [20,21], atomic force microscopy (AFM) [22,23] and scanning tunneling microscopy (STM) [24], have been used to image the surfactants near to the nanoparticle surface, but have been unable to achieve atomic resolution and/or measure the nanoparticle surface structure 'hidden' beneath the surfactant. Spectrometry techniques, such as small-angle neutron scattering, inductively coupled plasma-mass spectrometry, inductively coupled plasma atomic emission spectroscopy and X-ray photoelectron spectroscopy, have excellent chemical sensitivity [25–29]. However, their spatial resolution is limited to length scales (~10 nm to 10 micron) often larger than the nanocrystal itself, and certainly much larger than the facet or atomic dimensions. This has led to inconsistent reports in their application to detect metal additives on nanoparticle surfaces [13,25–28,30].

Conventional transmission electron microscopy (TEM) and scanning transmission electron microscopy (STEM) are powerful tools for imaging the atomic structure of nanocrystal surfaces [31–35]. However, achieving imaging conditions that are simultaneously sensitive to the metal nanocrystal and the surfactant is extremely difficult. Furthermore, the electron scattering processes, and hence image interpretation, at the nanoparticle-surfactant interface are severely complicated by the possibility of incomplete surface layers, trace amounts of metal additives and the structural discontinuity presented by the hard (nanocrystal)-soft (surfactant) interface. This means that several hypothetical models of the atomic structure of the nanoparticle-additive-surfactant interface matching the STEM or TEM images cannot be distinguished. These issues are further complicated by the extreme sensitivity to electron beam damage of the delicate surfactants and, to a lesser extent, the metal nanoparticle surfaces [36].

While all the above techniques provide vital information, the atomic structure of a nanoparticle-adsorbate-surfactant interface remains elusive.

In this work, we develop and apply an approach to determine the atomic structure and measure the interatomic distances at a nanoparticle-adsorbate-surfactant interface using STEM imaging conditions tuned specifically for this task by exploiting the specimen information encoded in the angular distribution of dynamically scattered electrons. Specifically, we use electron scattering calculations to identify imaging conditions that are differentially sensitive to the features of interest, namely, the nanocrystal surface facet, the metal-adatoms and the surfactant headgroup, permitting the generation of directly interpretable images of these features. Furthermore, we select those conditions that keep the electron dose relatively low to minimize damage. We apply these conditions to the gold nanocuboid system (grown in the presence of copper additives and CTAB) [11–14]. As noted above, $Cu^{2+}$ additives are essential to induce symmetry breaking and growth into a nanocuboid, as opposed to a nanocube or nanorod. It is thus important to resolve the location and atomic arrangement of Cu additives and their relationship to the gold surface and surfactant head group in order to provide insights into the underlying mechanisms by which additives induce anisotropic growth of metal nanocrystals. We acquire spatial maps of the angular distribution scattered from gold nanocuboids using a so-called Scanning Convergent Beam Electron Diffraction or 4D-STEM experimental configuration [37]. We harness the desired information to generate images sensitive to the interface, from which we generate an initial chemical model. By matching this model against the full angular range of the dataset, we measure the position and interatomic distances between the gold facet, copper adatoms and bromide at the surfactant headgroup.

**Results**

**Approximate location and distribution of metal additives and surfactants using conventional methods**

We first apply conventional methods to determine the approximate location and distribution of the metal additives and surfactants with low spatial resolution.

In conventional scanning transmission electron microscopy (STEM) imaging experiments, the electron probe is raster scanned across the specimen and at each position of the probe (x,y) and the scattered electrons are collected and summed over a fixed angular range using disc or annular detectors to form STEM Bright Field (BF) or Annular Dark Field (ADF) images, respectively. At the same time, x-rays emitted by the specimen can be detected with an x-ray spectrometer to deliver a map of chemical elements versus probe position (x,y) (Fig. 1).

We used energy dispersive X-ray spectroscopy (EDX) to obtain nanometre resolution chemical information regarding the distribution of copper additives and surfactants on the gold nanocuboid. We deliberately use a low spatial resolution of ~ 1 nm to keep radiation dose, and hence damage, relatively low (Fig. 1A). EDX reveals Cu on all the surfaces of the Au nanocuboid, although the distribution is not always even (Fig. 1A and Supplementary Note 1). Trace amounts of bromide near the surface are also detected, but we emphasize this signal is just marginally above the noise in the EDX spectra (bromide is expected because it is the counterion of the cetyltrimethylammonium ($CTA^+$) surfactant). In the vertically averaged EDX intensity across the nanocuboid in Fig. S2, it is evident that the Cu and Br are located somewhere within a ~2 nm wide region at the nanoparticle-surfactant interface. However, the resolution of this experiment does not permit a more precise assessment of the location. Obtaining EDX data at higher spatial resolution would require significantly higher electron dose and with this comes ambiguity arising from beam damage, including the degradation of surfactants, reconstruction of surface structure and alloying of Au and Cu, which are discussed in detail in Supplementary Note 2.

Having identified Cu additives and Br from the CTAB surfactant in the vicinity of the surface, we consider the location of the organic tail group ($CTA^+$) of CTAB. The qualitative distribution of surfactant is identified using conventional STEM-HAADF (Fig. 1B top, and Supplementary Note 3). The surfactant is distributed at a similar density on all facets and in this projected image extends about 1.5 nm from the nanoparticle {100} facets and about 1 nm from the corner facets. The projected surfactant length in these images is smaller than the 2-4 nm chain length of CTAB determined by neutron scattering [29] and STEM electron energy loss spectroscopy [38] and recently by liquid-cell TEM [39]. This may be because the carbon chains have collapsed and/or condensed while preparing the TEM sample, or because the surfactants are not aligned exactly perpendicular to the Au surfaces, and so appear shorter in projection [39].

The individual surfactants can just be resolved by examining the nanoparticle in the <110> orientation in STEM-BF (Fig. 1B bottom). In this projection, the surfactants attached to the thin end of the wedge (where two {100} facets meet) can be isolated. The surfactants are measured to be ~1.84 nm long, closer to the length of the CTAB molecule. As noted above, the cetyl chain may be curved, making the surfactant appear shorter in projection.

In principle, the conventional STEM techniques above can be applied at atomic resolution and may show atomic layers at the surface with different intensity. However, there are too many parameters influencing the intensity to correlate it with a unique structural model, as noted above and discussed in detail in Supplementary Note 4. New imaging modalities at doses lower than that needed for spectroscopic imaging must be developed to determine unambiguously the atomic model of the gold nanocuboid-adatom-surfactant surface.

**Identifying conditions to image the position of Cu relative to Au**

In conventional STEM, the integration of the scattered signal over a large angular range reduces the information about the material transferred to the final image. Recently, the advent of relatively fast, pixel array detectors has made it possible to record the full angular distribution of scattered electrons ($k_x, k_y$) at each probe position (x,y), so-called 4D-STEM (Fig. 1C).

From the above preliminary conventional experiments, we know that there are copper atoms, bromide ions and the electrostatically attracted CTA$^+$ cations in the vicinity of the surface of the Au nanocuboid. Though we cannot identify exactly where they are from those experiments, we can make a few reasonable starting assumptions. Firstly, from chemical considerations, we expect the copper atoms to be on the Au nanoparticle surface or alloyed within the particle. Secondly, based on the molecular dynamics calculations that suggest that bromide ions are present as the counterion of the surfactant (CTAB) [40], we might expect the bromide ions to adsorb at the nanoparticle surface, resulting in a negatively charged surface which the ammonium group of the CTA$^+$ cation would associate with. We now consider whether we can identify exact locations of these atoms using the additional information embedded in 4D-STEM datasets and thereby reconstruct an atomic model of the nanocuboid surface-adatom-surfactant interface.

We start with multi-slice simulations [41] of the electron scattering processes [42] to identify experimental conditions for distinguishing copper from gold at atomic resolution while minimising electron dose (Supplementary Note 5). From these simulations we identify a combination of experimental parameters particularly suited to achieving high differential sensitivity to copper while maintaining a relatively low electron dose. Specifically, we identify a relatively small incident probe convergence angle (9 mrad), and hence electron dose (compared with typical atomic resolution STEM), and scattering angles that reveal significantly different scattering behaviours between copper and gold atoms. Moreover, such behaviours persist across a wide range of possible specimen thicknesses, removing potential ambiguities in image interpretation associated with different atomic layer thicknesses, including incomplete surface layers.

To show the effectiveness of these conditions to provide sensitivity to the presence of Cu on the nanoparticle surface when they are implemented in 4D-STEM, we calculate and compare the 4D-STEM data set generated by two different idealized nanocuboid surface structures. Fig. 2 compares diffraction patterns across the interface for two model structures: one for pure Au (Fig. 2A) and the other where the two outer layers are instead Cu columns (Fig. 2B) (the latter as suggested by conventional STEM in Fig. S8 and Fig. S9). For each model, a series of diffraction patterns was simulated, with a 9 mrad probe scanning across the nanocuboid, from bulk into vacuum, in the direction perpendicular to a {100} surface facet.

For both nanocuboid models, the location of the nanocuboid-vacuum interface is evident in the loss of a vertical mirror plane in the diffraction pattern (most evident at position 5 in Fig. 2) parallel to the {100} face. At this discontinuity of the electron-specimen potential field, there is a stronger scattering in the <100> direction compared with the <$\bar{1}$00> direction, particularly just outside the central disc, as previously observed in scanning diffraction experiments [43,44]. We will call this the "discontinuity signal". Importantly, this asymmetric "flare" just beyond the central disc only exists in the direction perpendicular to the edge (regions marked by yellow boxes), enabling its differentiation from effects arising from surface atomic structure or composition.

When the probe is positioned within the pure Au nanocuboid, the STEM diffraction patterns are essentially identical to those of a pure Au crystal lattice (position 1 – 4, Fig. 2A). When the outermost two layers of the Au nanocuboid are replaced by Cu atoms and the probe is atop the Cu columns, the scattered intensity just beyond the central disc is significantly stronger than for pure Au (Fig. 2B), consistent with the predictions in Supplementary Note 5. This suggests a mechanism by which Cu may be distinguished from Au (or surfactant /vacuum) at the nanocuboid surface. In particular, if a detector selects and integrates the region of the diffraction pattern that is sensitive to copper (blue boxes in Fig. 2) but specifically excludes the region that is also sensitive to the structural discontinuity at the nanocuboid-vacuum interface (yellow boxes in Fig. 2), copper columns would appear much brighter than Au columns in the reconstructed image, independent of the presence of this interface. Importantly, simulations at different thicknesses (Fig. S12 and S13) indicate that this difference in intensity holds for a wide range of thicknesses, including the full range of possible thicknesses of these nanostructures and their surface layers, with no contrast reversal.

While many different types of images can be reconstructed from the 4D-STEM dataset (e.g. ptychographic, differential phase contrast and centre of mass), the proposed image reconstruction is tailored specifically to the problem at hand, generating images that can be interpreted in terms of Au or Cu atoms by direct inspection. This is only possible because of our a priori knowledge that these are the only elements present on or within the nanoparticle but it nevertheless provides an unequivocal approach for generating an image of the location of these two elements in the nanoparticle.

**Qualitative atomic model using copper-sensitive imaging conditions**

Using these parameters, 4D-STEM datasets were collected by a fast pixel array detector [45]. The position-averaged diffraction pattern (PACBED) shows that the nanocuboid is precisely oriented in the <100> zone axis (Fig. 3A). The total electron dose using these parameters for the 4D-STEM dataset, including the relatively small probe forming aperture ($5.6 \times 10^5$ e/Å$^2$) is much lower than the beam damage threshold estimated in Supplementary Note 2 ($9.9 \times 10^5$ e/Å$^2$). Moreover, this is 4 orders of magnitude lower than that used for the EDX experiments for the detecting of surface adatoms in Fig. 1A ($1.7 \times 10^9$ e/Å$^2$). Several images were reconstructed from this data set using different parts of the angular distribution of intensity, guided by the multi-slice calculations above.

The first reconstructed image is generated by integrating the regions of the diffraction pattern identified above as preferentially sensitive to the presence of copper while simultaneously excluding any complicating signal arising from the discontinuity at the nanoparticle-surfactant interface (green regions in Fig. 3B). In this way we can be confident the higher intensity signal evident at the surface in the reconstructed image (Fig. 3C) is due to the presence of copper and not complicated by the discontinuity signal (Fig. 3D).

To improve the signal to noise ratio and the resolution, a second image is reconstructed that uses all of the signals that are preferentially sensitive to copper, not just half of it (Fig. 3E, 3F). While the additional 2 detector quadrants (Fig. 3D by integrating C, D in Fig. 3B) will include some discontinuity signal, a comparison of the two copper-sensitive images with and without this signal shows a consistent form of the intensity distribution and location of the maxima, suggesting the copper signal dominates over the discontinuity signal (Fig. 3C, 3D, 3F). This is consistent with the full PACBED pattern in Fig. 3A, which only shows a very slight asymmetry between the intensity in quadrant C and D. (Note, we do not expect the discontinuity signal to be as strong as in the calculations in Fig. 2 because of the presence of the surfactant at the real interface, as opposed to the vacuum used in the idealised calculations.)

In the copper-sensitive image in Fig. 3F, from left to right across the nanocuboid, we observe low intensity atomic contrast in the bulk of the gold nanocuboid, a high intensity signal at the surface (red arrow), plus a monolayer of mid-intensity maxima at a nominal distance from the surface that is significantly larger than the gold lattice spacing (blue arrow). The superior resolution of the image using the full annulus also suggests the copper distribution comprises a complete monolayer followed by a half-occupied sublayer. We formulate an initial model by ascribing these intensity maxima to Au, Cu and Br (from left to right, Fig. 3G). At this stage, this model is the most plausible, informed by the aforementioned conventional EDX and STEM experiments in Fig. 1 and the tailored 4D-STEM experiments in Fig. 3F, plus the molecular dynamics calculations previously reported [40].

Note that additional images can be reconstructed from the 4D-STEM dataset which also reveal similar intensity features, such as by using non-negative matrix factorization (Supplementary Note 8). While these are reassuring, they are not generated with any prior understanding of the origin of these features, unlike the directly interpretable tailored image in Fig. 3C, and therefore they are not used here to infer an atomic model.

Also note that we also considered other mechanisms (other than different atomic species) that might cause bright surface contrast, for example the relaxation of surface atoms or an increase in thermal vibrations of surface atoms, but these were found to be unlikely (Supplementary Note 9).

**Quantitative atomic structure with interatomic distances**

The above has identified the location of copper and provided a qualitative structural model for the Au nanocuboid-additives-surfactant (Au-Cu-Br) interface (Fig. 3G). To further validate and improve the model, we use the known structure of the interior of the Au nanocuboid as a reference to quantify various experimental parameters, which enables the measurement of nanocuboid thickness profile and the interatomic distances of the structural model.

We first measure the thickness variation approaching the nanocuboid surface. Averaged experimental diffraction patterns from gold columns are compared to simulated diffraction patterns at different thicknesses by a template matching method (Fig. S20). Through the thickness determination of all gold atomic columns from bulk towards the surface, we reconstruct the thickness profile of the nanocuboid (Fig. 4A).

Knowledge of this thickness profile removes significant variables and uncertainties from image simulations and opens the possibility of quantitative measurement of the interatomic distances in our, hitherto, qualitative structural model of the nanocuboid surface-surfactant interface (Fig. 3G). To do this, we simulate the 4D-STEM data for this thickness-refined structural model and quantitatively compare it with the experimental dataset across different angular ranges. We adjust and refine the positions of surface Cu and Br layers to obtain a match (Fig. 4B to 4D). In this process, the large variety of different angular ranges is important to constrain the solution by ensuring the number of unknowns is much smaller than the knowns (Supplementary Note 10). This leads to refined atomic distances in projection of 1.81 Å from the last Au layer to the first Cu layer, 1.13 Å from the first Cu monolayer to the second Cu sublayer and 2.56 Å from the Cu sublayer to the Br surfactant head group layer. As a reference, the distance between Au layers are 2.04 Å. The resultant structure model is shown in Fig. 4E.

The degree of validity of this quantitative model and interatomic distances is displayed in the comparison of simulated and experimental data in Fig. 4 (B-D). The contrast at the nanocuboid surface observed in all four sets of images exhibits close agreement between the simulated and experimental data, despite the image streaking caused by scan 'noise' and specimen movement (due to the relatively slow acquisition time for this first-generation pixel array detector). Moreover, the intensity profiles accurately match the simulation results, in terms of peak positions and relative peak intensities, indicating that the final quantitative structure model, including the projected interatomic distances, reliably represents the structure of the Au nanocuboid-surfactant interface.

**Discussion**

This work demonstrates an approach to determine the atomic structure at the hard-soft interface of a metal nanoparticle grown in the presence of metal adatoms and surfactants, which has hitherto been elusive. The approach starts by using conventional methods to measure the surface profile and composition at low resolution. Using this preliminary information plus dynamical electron scattering calculations, we purpose-design an atomic resolution imaging mode that can be interpreted directly in terms of metal additives and gold atoms, while minimizing uncertainties from thickness variation, partially occupied atomic layers and/or nanoparticle-surfactant interfaces. Furthermore, it does this at an undamaging electron dose. From these purpose-designed images, we identify the location of metal additives, in this case copper, on the surface without ambiguity and generate a qualitative atomic model for the Au-metal additive-surfactant interface. We then validate and quantify this model using the full 4D-STEM data, which heavily constrains the solution.

This approach allows for the unambiguous determination of the Au-Cu-Br surface atomic structure, including determination of the nanoparticle thickness profile and, critically, measurement of the projected interatomic distances between Au, Cu and Br at the interface.

Notably, we reveal that Cu additives form a fully-occupied monolayer on the gold nanocuboid surface, followed by a half-occupied second layer. Cu ad-layers are followed by the surfactant counterion (Br$^-$) at a distance of 2.56 Å to the outer most Cu layer, confirming their complex interaction with the additives. The organic CTA$^+$ ion head group interacts presumably electrostatically with the Br$^-$ at the surface. This provides direct evidence for nanoparticle growth mechanisms which propose control of nanoparticle growth via underpotential deposition of metal additives [46,47]. It is the complex nature of this surface which leads to symmetry breaking and the formation of the single-crystal nanocuboid products.

The 4D-STEM approach presented herein demonstrates how the atomic structure of complex hard-soft interfaces can be discerned, taking advantage of the rich information encoded in the full diffraction patterns at each probe position. We show how it overcomes the limitations of conventional STEM and STEM-EDX by concurrently enhancing sensitivity and minimizing beam damage. This approach holds immense promise for exploring and defining nanocrystal surface interactions with additives, surfactants, and the surrounding environments.

**Methods**

*Chemicals*

Gold (III) chloride trihydrate (HAuCl$_4$·3H$_2$O) (≥ 99.9%), copper (II) chloride (CuCl$_2$) (≥ 99.999%), sodium borohydride (NaBH$_4$) (≥ 99.9%), and L-ascorbic acid (≥ 99.9%) were purchased from Sigma-Aldrich. Hexadecyltrimethylammonium bromide (CTAB) (98%) was purchased from Ajax Finechem. Ultrapure water (milli-Q, $R$ > 18.2 MΩ·cm) was used for the preparation of all solutions. All glassware used for the syntheses were cleaned using aqua regia prior to use.

*Gold nanocuboid synthesis*

**Seeds:** Aqueous solutions of CTAB (1.875 mL, 0.2 M), HAuCl$_4$ (0.025 mL, 0.05 M) and milli-Q water (2.83 mL) were mixed thoroughly. Aqueous NaBH$_4$ solution (0.3 mL, 0.01 M) was then rapidly added into the solution with vigorous stirring, and stirred for an additional 10 min. The resulting seed solution is kept at 28 ± 1 °C for 40 min prior to use. The seed solution was diluted 10 times with milli-Q water, immediately prior to the further use for growth of gold nanocuboid.

**Growth:** Aqueous solutions of CTAB (1 mL, 0.2 M) and HAuCl$_4$ (0.1 mL, 0.05 M) were mixed with milli-Q water (11.38 mL). The resultant solution was thoroughly stirred and left for at least 5 min to ensure homogenisation. Aqueous ascorbic acid solution (1.5 mL, 0.1 M) was then added, mixed thoroughly, and after 5 minutes, aqueous CuCl$_2$ (27 μL, 0.01 M) was added, and mixed thoroughly. After 1 minute, the diluted seed solution was added and mixed thoroughly. The nanocuboids are grown for 48 hours, at 28 ± 1 °C. The nanocuboids were purified via centrifugation and redispersion to give a final concentration of CTAB of 0.001 M.

*TEM experiments*

The ultrathin carbon coated Cu TEM grids were plasma cleaned in H$_2$/O$_2$ for 30 seconds before use. A sample of the cleaned nanocuboids was immediately dropped onto the TEM grid and left for 5 min. The grid was placed into ultra-pure ethanol for 20 min to remove excess CTAB. Plasma cleaning was avoided to retain the pristine CTAB attached to the nanoparticles.

Electron microscopy was carried out at the Monash Centre for Electron Microscopy (MCEM) on a FEI Titan$^3$ 80-300 kV FEGTEM and a Thermo Fisher Scientific Spectra ɸ FEGTEM equipped with both probe and imaging spherical aberration correctors for (S)TEM. Conventional STEM experiments were carried out at 300 kV with a 15 mrad probe-forming aperture and 4D-STEM experiments with a 9 mrad probe-forming aperture. 4D-STEM datasets were recorded with a hybrid pixel array detector (electron microscopy pixel array detector, or EMPAD) at 1 kHz framing rate.

*CBED simulations*

Images and diffraction patterns were calculated using a GPU-enhanced frozen-phonon multi-slice code (μSTEM) [42], including multiple elastic scattering and multiple thermal diffuse scattering. An aberration-free probe was assumed. 30 passes were carried out to ensure the accuracy of the frozen-phonon calculations.

**Data availability**

All relevant data generated in this study are provided in the paper and its Supplementary Information files. Source data has also been deposited in Figshare under the accession link xxx. Source data are provided with this paper.

**Acknowledgments**

The authors acknowledge Dr. Michael Walsh and Dr. Dan Nguyen for helpful discussion and Dr. Matthew Weyland and Dr. Matus Krajnak for assistance with 4D-STEM experiments.

**Funding**

We acknowledge gratefully funding from the Australian Research Council (ARC), including ARC Discovery Project DP160104679 (J.E. and A.F.), ARC Centre of Excellence in Exciton Science, CE170100026 (A.F.), ARC Laureate Fellowship FL220100202 (J.E) and ARC Future Fellowship FT190100619 (S.F). W.L. thanks the support of an Australian Government Research Training Program (RTP). Work at the Molecular Foundry was supported by the Office of Science, Office of Basic Energy Sciences, of the U.S. Department of Energy under Contract No. DE-AC02-05CH11231. The authors acknowledge use of facilities within the Monash Centre for Electron Microscopy, a node of Microscopy Australia. The Thermo Fisher Scientific Spectra φ TEM was funded by ARC LE170100118 and the FEI Titan$^3$ 80-300 FEG-TEM was funded by ARC LE0454166.

**Author contributions**

W.L., A.F. and J.E. conceived the idea and designed the experiments. W.L. carried out TEM experiments, simulations and analyzed the data. W.T. and A.Y synthesized gold nanoparticles under the supervision of A.F.. C.O. performed NNMF analysis. B.E., S.F. and T.P. participated in the TEM simulations and interpretation of experimental data. C.Z. assisted with preliminary TEM experiments. W.L. and J.E. wrote the manuscript with input from all co-authors.

**Competing interests**

The authors declare that they have no competing interests.

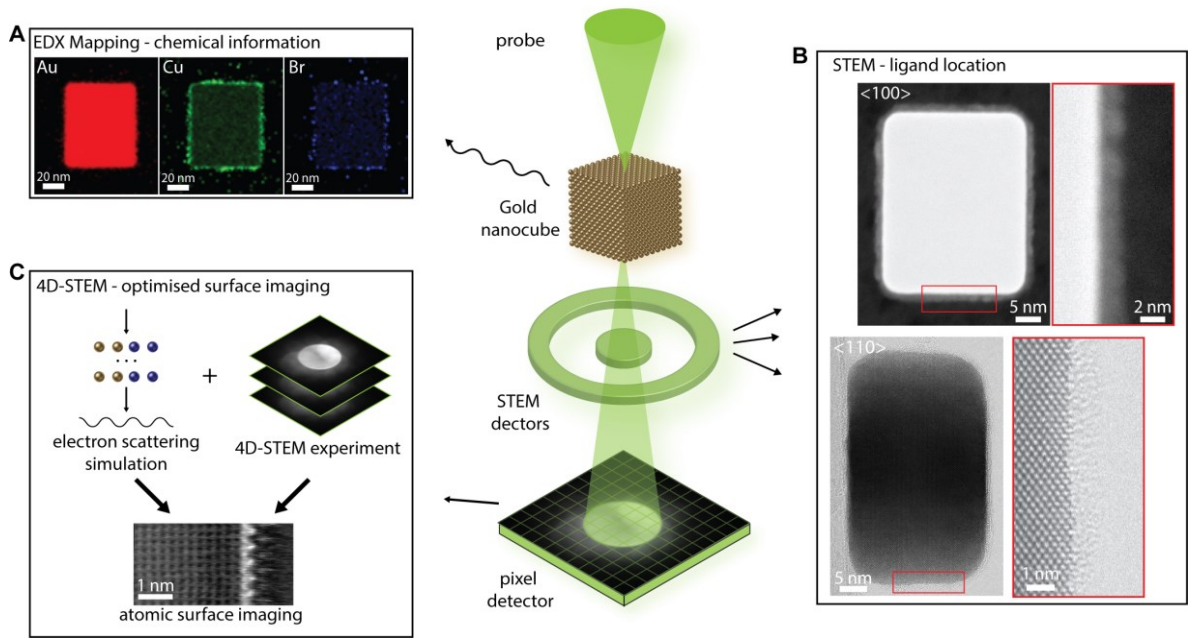

**Fig. 1. Overview of STEM experiments in characterizing copper additives and CTAB surfactants on a gold nanocuboid surface.** A focused electron probe is scanned across the nanocuboid surface, and at each probe position electron signals are collected by various detectors. **(A)** EDX maps provide the location of additives (Cu) and surfactant counterion (Br) at nanometer resolution. **(B)** Conventional STEM integrates the diffraction pattern using a disc or annular detector. Top: <100> zone axis STEM-HAADF image of surfactant attached to the {100} facets; Bottom: Atomic resolution <110> zone axis STEM-BF image of surfactant attached to the edge where two {100} facets meet. Red-boxed images on the right are the magnified, rotated images of the red box annotation in the images on the left. **(C)** 4D-STEM collects the diffraction pattern at each probe position with a pixel array detector. Electron scattering simulations identify conditions for direct imaging of the Au-Cu-surfactant interface. Under these conditions, a collection of diffraction patterns is collected and a direct image of the Au-Cu-surfactant interface is generated from which the atomic arrangement and distances will be measured.

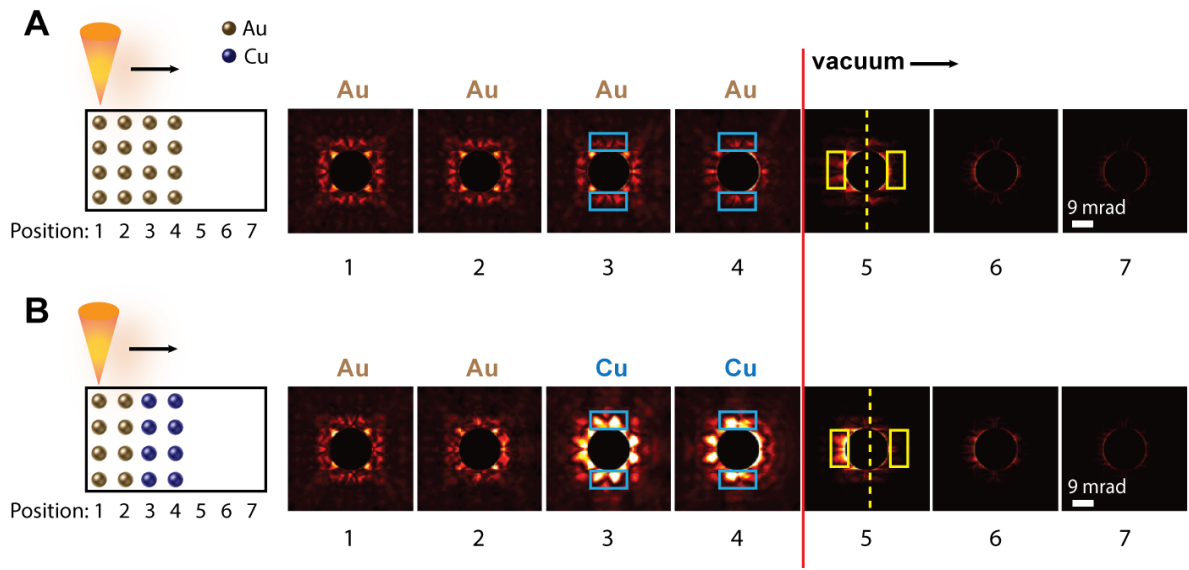

**Fig. 2. Simulation of convergent beam electron diffraction patterns with a 9 mrad convergence angle (corresponding to a 1.3 Å diameter probe) moving across a gold nanocuboid into vacuum (position 1-7). (A)** Pure gold nanocuboid. **(B)** Equivalent simulation with the last two layers replaced by Cu atoms. Model thickness is the average size of gold nanocuboids, 25 nm. The probe step size is 2.04 Å so that the probe positions coincide with the atomic column positions. Central discs are masked for better observation of the intensity distribution beyond the central disc (unmasked raw diffraction patterns are shown in Fig. S11). The colored intensity scale, common to all diffraction patterns, ranges from 0 to 0.005% of the maximal vacuum probe intensity. Blue boxes indicate the features that are sensitive to copper atoms but are insensitive to the vacuum interface. Yellow boxes indicate the features arising from the nanocuboid-vacuum interface. The interface breaks the mirror symmetry in the direction perpendicular to the surface, indicated by the differences in reference to the dashed yellow line.

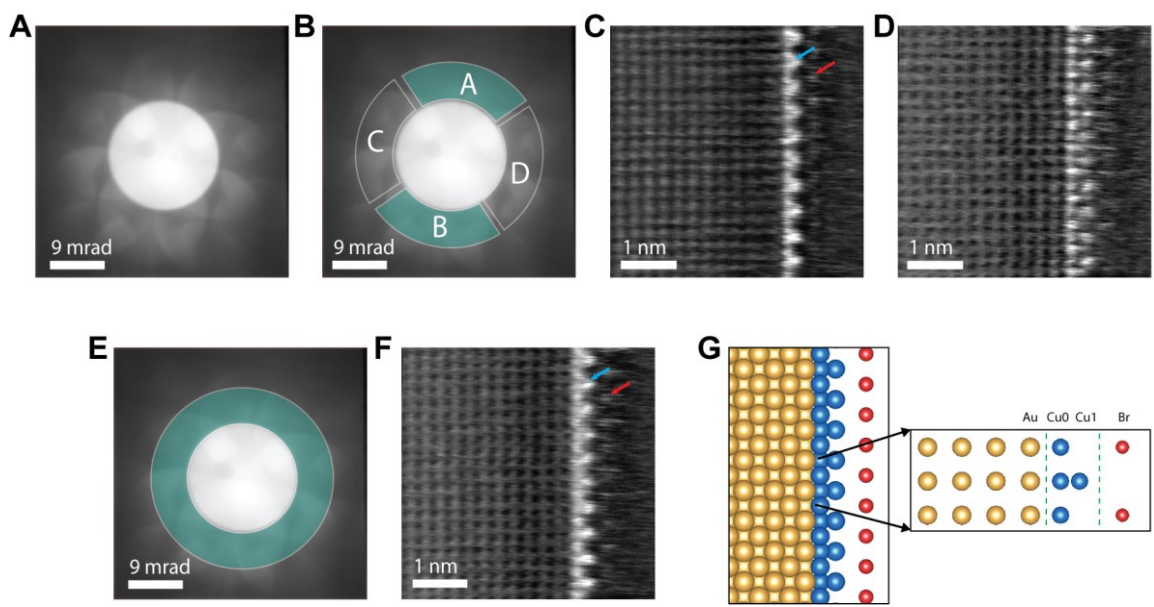

**Fig. 3. Experimental 4D-STEM dataset collected by scanning a 9 mrad probe across a gold nanocuboid surface. (A)** PACBED pattern. **(B)** Optimized copper sensitive detector determined by multi-slice simulations. Images are reconstructed by integrating segments labeled by **(C)** A+B, and **(D)** C+D. **(E, F)** Image reconstructed by integrating the whole 9-15 mrad annular range. **(G)** Schematic surface model derived from the reconstructed images (yellow atoms – Au, blue atoms – Cu0, Cu1, red atoms – Br). The leftmost dashed line is midway between the outermost Au atoms and the Cu0 layer, while the other line demarcates the boundary between the Br and Cu1 layer. The relative rotation of real space and diffraction pattern is calibrated in Supplementary Note 6. Reconstructed images are scan-artefact corrected and filtered as detailed in Supplementary Note 7. Arrows in **(C, F)** indicate surface adatoms: blue – Cu, red – Br.

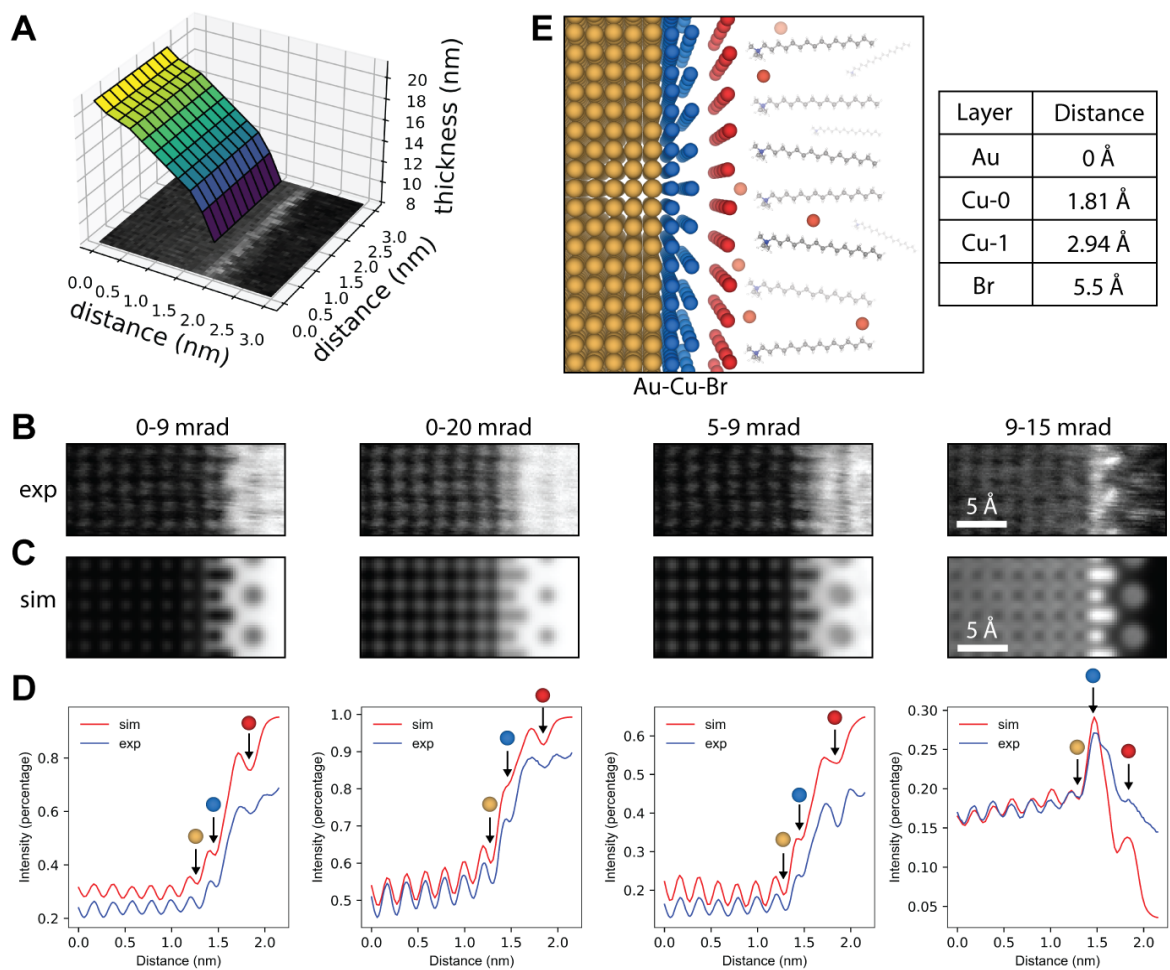

**Fig. 4. Refinement of the gold nanocuboid surface model. (A)** Reconstructed nanocuboid thickness plot close to the nanocuboid surface. **(B-D)** Validation of the final model: comparison of simulated 4D-STEM data of the atomic structure **(E)** with experiment. Images are reconstructed from the 4D-STEM data by integrating across 4 different angular ranges with a quantitative comparison of their vertically integrated intensity profiles. **(E)** Final, quantitative atomic structure of the Au nanocuboid-copper-bromide interface with measured interatomic distances (as projected in <100> direction) (Au - yellow, Cu - blue, Br - red): Au-Cu0=1.81Å, Cu0-Cu1=1.13Å, Cu1-Br=2.56Å.